# FuzzingDriver: the Missing Dictionary to Increase Code Coverage in Fuzzers

Arash Ale Ebrahim*, Mohammadreza Hazhirpasand‡, Oscar Nierstrasz‡, Mohammad Ghafari‖
* Independent security researcher, Singapore
‡ University of Bern, Switzerland
‖ University of Auckland, New Zealand

***Abstract*—We propose a tool, called FuzzingDriver, to generate dictionary tokens for coverage-based greybox fuzzers (CGF) from the codebase of any target program. FuzzingDriver does not add any overhead to the fuzzing job as it is run beforehand.**

**We compared FuzzingDriver to Google dictionaries by fuzzing six open-source targets, and we found that FuzzingDriver consistently achieves higher code coverage in all tests. We also executed eight benchmarks on FuzzBench to demonstrate how utilizing FuzzingDriver's dictionaries can outperform six widely-used CGF fuzzers.**

**In future work, investigating the impact of FuzzingDriver's dictionaries on improving bug coverage might prove important. Video demonstration : https://www.youtube.com/watch?v=Y8j_KvfRrI8**

## I. Introduction

Compared to non-security issues, only a small group of developers are involved in reporting security issues in software projects [1]. Fortunately, with the increasing adoption of modern fuzzers, developers have been able to efficiently find critical vulnerabilities. For instance, the AFL fuzzer, a coverage-based Greybox Fuzzer (CGF), has found a large number of security vulnerabilities in well-known software programs.[1] Researchers have also been actively working on improving coverage-based guided fuzzers. As an example, optimized mutation algorithms introduced by MOpt have had a great impact on increasing code coverage and discovering more vulnerabilities [2]. Similarly, AFLSmart introduced structure-aware fuzzing by combining the PEACH fuzzer engine with the AFL fuzzer, and the Angora fuzzer employs data-flow analysis, yielding a noticeable improvement in the level of code coverage [3, 4].

Most of the coverage-guided fuzzers, *e.g.,* AFL and Libfuzzer, make use of dictionaries to optimize the fuzzing job and induce more interesting paths. Dictionaries are generally flat ASCII files where tokens, extracted strings from the target program, are listed per line. Each line can consist of *key:value* pairs, where the key is optional. By utilizing dictionaries, fuzzers exercise relevant tokens, which significantly improves the likelihood of finding new paths and use cases [5]. However, dictionary generation is considered to be an arduous and time-consuming task, and users often simply reuse the predefined dictionaries introduced by the AFL fuzzer or Google. Unfortunately, such dictionaries are general-purpose and lack target-dependent specifications. This limitation in real-life scenarios leads to executing fuzzers without a relevant dictionary and consequently obtaining less code coverage or increasing the time and cost budget.

We introduce a tool, named FuzzingDriver, to automatically generate dictionaries for each program. FuzzingDriver utilizes CodeQL to extract valuable information, [2] *i.e.,* commonly occurring keywords, strings and constants, from the internals of the target program. Such information assists fuzzers to mutate the expected values that reside in various branches, *e.g.,* checksums, and achieve a higher code coverage compared to when they employ general-purpose dictionaries. FuzzingDriver generates dictionaries related to various file formats or protocols, and can be employed as a plug-in for CGFs.[3]

We evaluated FuzzingDriver in two different experiments. We first used FuzzingDriver and Google dictionaries for fuzzing six open-source binaries. In all six binaries, FuzzingDriver helped to achieve higher code coverage. In the second experiment, we executed eight benchmarks on FuzzBench in order to evaluate the efficacy of the AFL fuzzer armed with FuzzingDriver's dictionaries against six state of the art CGF fuzzers. We achieved promising results as FuzzingDriver's dictionaries significantly increased the code coverage of AFL. In particular, we obtained the highest code coverage in six binaries and substantially higher coverage than the rivals in three binaries. Assessing the effectiveness of FuzzingDriver in terms of bug coverage as well as its optimization to minimize overhead can constitute the object of future studies.

The remainder of this paper is structured as follows. In section II, we explain how customized dictionaries improve the fuzzing job. We present our workflow in section III and discuss our results in section IV. We discuss related work in section V. We conclude the paper in section VI.

## II. Why customized dictionaries?

Fuzzers commonly do not offer any automated approach to generate dictionaries from target programs. Therefore, it is of interest to propose a tool to accomplish this task with minimum effort and observe to what extent FuzzingDriver's cus-tomized dictionaries can help the performance of (mutation-based) CGFs, such as AFL,[4] AFLFast [6], Steelix [7], VUzzer [8], and Angora [4]. In the following, we demonstrate the advantage of employing FuzzingDriver's dictionaries in fuzzing.

[1] https://lcamtuf.coredump.cx/afl/

[2] https://securitylab.github.com/tools/codeql/

[3] https://github.com/cryptomadco/fuzzingdriver

We elucidate a small yet complex example, shown in Listing 1, to pinpoint a problem with state of the art fuzzers.[5] The code reads an input file and passes the content of the file, *e.g.,* data, and the size of the content, as arguments to a user-defined function called *Test*. In the Test function, there are complex branches in which conditions must be satisfied in order to set each flag to an integer number, which is the corresponding branch number. If all the flags are set, the final branch will trigger a violation, and accordingly, the fuzzer will mark it as a bug.

We used a machine powered by Ubuntu 18.04 x64, 32 CPU cores with 64GB of RAM for two experiments. Each test is repeated five times and each time has a duration of about five hours. In the first experiment, we selected QSYM[9] to execute hybrid fuzzing (fuzzing and concolic execution) in order to traverse the difficult branches. Prior to this decision, we also selected other CGFs like AFL and libfuzzer but they failed as the example includes many complex branches. QSYM discovered the vulnerable point after 4 hours and 54 minutes on average.

Listing 1. The motivational example–partial code snippet

```
uint8_t flags[8] = {0};
int Test(const uint8_t *Data, size_t Size) {
   uint32_t *num = (uint32_t *)Data;
   if (num[0] > 0x003e9ef4 && num[0] < 0x00649689) {
      if (num[1] > 0x00b10797 && num[1] < 0x00f2deeb) {
         if ((num[0] * num[1]) == 0x00621a27 * 0x00c01752) {
            flags[0] = 0;
         }
      }
   }
// More similar complex constraints *****
   if (num[14] > 0x073f66a5 && num[14] < 0x07f04124) {
      if (num[15] > 0x07414558 && num[15] < 0x078e3e98) {
         if ((num[14] * num[15]) == 0x074fd355 * 0x075e1841) {
            flags[7] = 7;
         }
      }
   }
#if 0
#endif
   if (flags[0] == 0 && flags[1] == 1 && flags[2] == 2 && flags[3] == 3
   && flags[4] == 4
   && flags[5] == 5 && flags[6] == 6 && flags[7] == 7 ) {
      *((volatile uint8_t *)0) = 0;
   }
   return 0;
}
```

In the second experiment, FuzzingDriver extracted magic numbers from the motivational example and generated a dictionary of useful tokens prior to the fuzzing job. We then executed the QSYM fuzzer with the dictionary generated by FuzzingDriver. Interestingly, QSYM discovered the vulnerable point noticeably faster in just 26 minutes on average. This suggests that effective dictionary generation can substantially help fuzzers in quickly discovering new paths in the target applications.

## III. The Workflow

FuzzingDriver comprises two parts: (1) extracting valuable tokens from the target's source code, and (2) a data cleaner script written in Python.

[4]http://lcamtuf.coredump.cx/afl/

[5]The complete source code of the example

*1) Literal extraction:* We prepared eleven CodeQL queries to extract valuable information, *i.e.,* tokens, from the codebase. CodeQL is an industry-leading semantic code analysis engine introduced by GitHub.
The flexibility of CodeQL enables the users of FuzzingDriver to write their own custom queries or modify the existing queries to expand FuzzingDriver's functionality. With regards to token extraction queries, we wrote one query for finding literals, one query for identifying user-defined comparison functions, two queries for arrays and global variables, and seven queries for looking into the arguments to the following functions: strstr(), strcasecmp(), strncasecmp(), strcmp(), strncmp(), memcmp(). These last functions are considered critical for fuzzers in order to satisfy various branches [10]. For instance, the importance of covering such functions can be viewed in the FFmpeg program in which there are various checks that are dependent on the strcmp function (See Listing 2).

Listing 2. The usage of strcmp() in the FFmpeg program

```
if (!strcmp(key, "acodec")) opt_audio_codec (o, key, value);
     else if (!strcmp(key, "vcodec")) opt_video_codec (o, key, value);
     else if (!strcmp(key, "scodec")) opt_subtitle_codec(o, key, value);
     else if (!strcmp(key, "dcodec")) opt_data_codec (o, key, value);
```

We also extract literals as hexadecimal values that are present in the program (See Listing 4). These values commonly tend to be part of specific checksums or important parts of the program. Listing 3 shows the significance of capturing the hexadecimal values in the libxml library.

Listing 3. Hexadecimal values in the libxml library

```
else if (((*in >= 0x20) && (*in < 0x80)) ||
    (*in == '\n') || (*in == '\t')) {
      // some code
   } else if (*in >= 0x80) {
      if (outend - out < 11) break;
      if (*in < 0xC0) {
      // some code
      } else if (*in < 0xE0) {
      if (inend - in < 2) break;
      val = (in[0]) & 0x1F;
      // some code
      }
```

Listing 4. Capturing literals

```
HexOrOctLiteral(){
    (this instanceof HexLiteral) or (this instanceof OctalLiteral)
   }
```

If a developer implements a custom string comparison function within the language, then this could prevent FuzzingDriver from detecting useful tokens. For instance, there are some custom functions such as xmlStrcasecmp and g_ascii_strncasecmp in the Libxml and glib libraries respectively that they in fact rely on the Strcasecmp function. To resolve this issue, we wrote a CodeQL query containing regexpMatch() that captures comparisons in the target program when a user-defined comparison function is called and tracks how parameters flow by the *DataFlow* module (See Listing 5). The regexpMatch() function considers every function in the program whose name contains the following strings: *str, mem, strn, cmp*. The list can be extended to cover a greater range of keywords in the regex section. However, it may not detect user-defined functions that have completely arbitrary names. This is noteworthy that the other CodeQL queries can be enhanced by using the *DataFlow* module to track how data flow through the parameters.

Listing 5. Capturing comparison custom defined functions

```
CmpArgNode() {
   exists(FunctionCall fc |
     fc.getTarget().getName().regexpMatch(".*(str|mem|strn|b)*(cmp|str)*") and
     fc.getArgument(0) = this.asExpr()
   )
```

As shown in Listing 6, to capture static constant values, the CodeQL query captures values such as the one demonstrated in Listing 7. The source code of the remaining queries is available online.[6]

Listing 6. Capturing static constants

```
import cpp
from GlobalVariable a
where a.isStatic() and a.isConst()
select a.getAnAssignedValue().getValueText()
```

Listing 7. A static constant holding valuable information

```
static const int extend_test[16] = { /* entry n is 2**(n-1) */
  0, 0x0001, 0x0002, 0x0004, 0x0008, 0x0010, 0x0020, 0x0040, 0x0080,
 0x0100, 0x0200, 0x0400, 0x0800, 0x1000, 0x2000, 0x4000
};
```

*2) Data cleaner:* It is essential to have a dictionary containing only useful tokens from the target program to reduce the fuzzing overhead. The main role of FuzzingDriver's data cleaner is to examine the extracted tokens and remove noisy tokens based on different criteria. In programs, there exist strings, *e.g.,* warning or error messages, that certainly do not expedite the process of fuzzing and squander effort on mutating incorrect, worthless tokens. Therefore, we provide users with configurations so that they can define the minimum and maximum length of tokens. It is also possible to define an array of prohibited characters, *e.g.,* white space, to either remove the tokens containing such characters or replace the character with a predefined one. We enable users to explore statistics regarding the found English keywords and their distribution in the target. We used the Levenshtein distance [11] to measure the distance between two tokens. The users are able to adjust the desired value for the distance between two tokens. As a result, they can observe a list of duplicated tokens and the ones whose distance is equal to or below the defined Levenshtein threshold. All these features allow the users to customize the output dictionary so that it contains only useful tokens and is free of ineffective ones.

## IV. Evaluation and Results

We evaluated how FuzzingDriver can improve the efficacy of CGF-based fuzzers. To this end, we tested our proposed technique against two other approaches and compared the obtained code coverage with the rivals. In the first approach, we used Google dictionaries, which became available to the public in May 2020.[7] Google provides a number of dictionaries for various file formats, *e.g.,* CSV and JS, which can be used by CGF-based fuzzers. In the second approach, we executed eight benchmarks on FuzzBench [12], Fuzzer Benchmarking As a Service. FuzzBench is a free service that allows researchers to rigorously assess their fuzzers on a wide range of real-world benchmarks (*i.e.,* open-source programs). We tested FuzzingDriver against six other state-of-the-art CGF fuzzers.

[6]https://github.com/cryptomadco/fuzzingdriver

[7]https://github.com/google/fuzzing/tree/master/dictionaries

TABLE I
THE OBTAINED BASIC BLOCK CODE COVERAGE BETWEEN GOOGLE DICTIONARIES AND FUZZINGDRIVER WITH SEVEN BINARIES

| Target | Dictionary | Avg | T1 | T2 | T3 | T4 | T5 |
|---|---|---|---|---|---|---|---|
| libjpeg | Google-dict | 1111 | 984 | 904 | 1128 | 983 | 1557 |
| | FuzzingDriver | 2553 | 4595 | 2020 | 1708 | 2422 | 2019 |
| poppler | Google-dict | 4246 | 4154 | 4093 | 4122 | 4380 | 4483 |
| | FuzzingDriver | 4388 | 4256 | 4331 | 4699 | 4112 | 4542 |
| binutils | Google-dict | - | - | - | - | - | - |
| | FuzzingDriver | 9370 | 9332 | 9612 | 9379 | 9522 | 9005 |
| libarchive | Google-dict | - | - | - | - | - | - |
| | FuzzingDriver | 10553 | 9939 | 11333 | 10882 | 10382 | 10231 |
| libxml | Google-dict | 12782 | 12764 | 12592 | 13036 | 12046 | 13473 |
| | FuzzingDriver | 13640 | 13829 | 13751 | 13390 | 13986 | 13246 |
| UnRAR | Google-dict | - | - | - | - | - | - |
| | FuzzingDriver | 9226 | 9955 | 9047 | 9036 | 9017 | 9073 |

### A. Google dictionaries

We selected six popular open-source libraries and software, namely Libjpeg, Poppler, libxml, binutils, libarchive, and UnRAR. The aforementioned binaries were used in various research works to assess the fuzzers' efficiency [6, 13]. Moreover, we chose the latest version to date (3.12c) of AFL++ [14] and instrumented all the targets with AFL++ SanitizerCoverage. For each test, we then fed the corresponding FuzzingDriver and Google dictionary to AFL++. The configuration of the machine running the fuzzing jobs was Ubuntu 18.04 x64, 32 CPU cores with 64GB of RAM.

Table I depicts the results of FuzzingDriver compared with Google dictionaries in each trial run (denoted by T), and on average of the five trial runs. FuzzingDriver outperformed Google dictionaries in all the binaries, whereas Google dictionaries did not provide any dictionaries for three targets. Google dictionaries can be helpful when standard specifications are expected while they lack some of the most well-known file formats. For example, Google does not present any dictionaries for file formats such as TAR and ELF concerning the libarchive and binutils binaries, respectively. On the other hand, FuzzingDriver substantially improved the fuzzing jobs in almost all of the trial tests and also on average due to careful dictionary generation and cleansing phases.

We used a tool named AFL-cov to visualize the coverage provided by FuzzingDriver.[8] AFL-cov employs test cases produced by the AFL-based fuzzers and interprets from one test case to the next one to determine which new functions and lines are hit by AFL. We selected some parts of the code which had been examined by using FuzzingDriver and Google dictionaries. Figure 2 shows the part of the code in the libxml binary that had been fully covered by the fuzzer with the help of FuzzingDriver, however the right side, indicated with the red color, shows that the same piece of code had not been traversed when Google dictionaries were used. We also realized that FuzzingDriver performed well on targets in which encoding/decoding occurs frequently. As another example, we witnessed that the function responsible for carrying out MCU decoding on libjpeg is fully covered with the help of the tokens extracted by FuzzingDriver while AFL++ ould not traverse the same path with the help of Google dictionaries.

[8]https://github.com/vanhauser-thc/afl-cov

### B. FuzzBench

The reason that we selected the FuzzBench platform is the fact that the platform was mentioned for benchmarking fuzzers in a number of recent research work [14, 15]. We ran FuzzBench [12] to conduct a benchmark between the AFL fuzzer armed with FuzzingDriver's dictionaries and other state-of-the-art fuzzers on FuzzBench. More precisely, we evaluated the AFL enabled FuzzingDriver fork against the original AFL, Fairfuzz[16], MOpt[2], AFLSmart[3], AFLFast[6] and Lafintel fuzzers.[9] We conducted our experiment on eight targets with five trial runs and each trial duration was ten hours. As shown in Figure 1, FuzzingDriver significantly improved the code coverage of the AFL in all the benchmarks. In benchmark with five binaries, FuzzingDriver gained the highest code coverage among the tested fuzzers. Moreover, in the rest of the binaries, FuzzingDriver achieved second place among its competitors. Comparing the original AFL with the AFL armed with FuzzingDriver reveals that the boosted version with FuzzingDriver performed substantially better in four targets, namely libpcap, libxml, libpng, and proj4. FuzzingDriver distinctly gained more coverage in three binaries, namely libpcap, libpng, and libxml. However, in three binaries (*i.e.,* bloaty, php parser, and libjpeg), the performance of the fuzzers is somewhat similar. To observe a detailed version FuzzingDriver's trial tests, please visit the fuzzer's repository.

## V. Related work

Wang *et al.* proposed a novel data-driven seed generation approach, called Skyfire, that processes a large number of existing samples, extracts the knowledge of grammar and semantic rules, and generates well-distributed seed inputs for fuzzing programs [17]. However, Skyfire is designed to work with a probabilistic context-sensitive grammar (PCSG) to specify both syntax features and semantic rules and its purpose is different from the current study. Householder and Foote proposed a workflow for black-box fuzzing and an algorithm for selecting parameters to maximize the number of unique errors detected in a fuzzing campaign [18]. AFLFast strengthened its successor's performance, AFL, by various strategies to exercise the low-frequency paths. AFLFast discovered three previously unreported CVEs that are not discovered by AFL and identified 6 previously unreported CVEs 7x faster than AFL [6]. Li *et al.* proposed GTFuzz that is capable of seed input prioritization, dictionary generation and offers an enhanced seed mutation algorithm [19]. However, they did not provide any source code of their work and we could not make any comparison. Similar to FuzzingDriver, Mathis *et al.* proposed a new approach, called LFuzzer, to automatically extracts tokens using dynamic tainting of implicit data transformations [20]. However, our approach necessitates less complexity to understand and is easier to extend.

[9] https://lafintel.wordpress.com/

## VI. Conclusion

We presented a tool, called FuzzingDriver, for dictionary generation for any targets. It identifies tokens from any file formats that are necessary to traverse a program. We evaluated FuzzingDriver against Google dictionaries and showed that FuzzingDriver significantly increases code coverage in fuzzing. We also demonstrated that FuzzingDriver helps the AFL fuzzer to outperform six widely-used CGF fuzzers in terms of code coverage. Further research is needed to investigate the impact of FuzzingDriver on bug coverage.

## VII. Acknowledgments

We gratefully acknowledge the financial support of the Swiss National Science Foundation for the project "Agile Software Assistance" (SNSF project No. 200020-181973, Feb. 1, 2019 - April 30, 2022). We are also grateful to Jonathan Metzman and Abhishek Arya from the Google OSS-FUZZ project, Marc Heuse from the AFL++ Project, and Stefan Nagy from Virginia Tech University for their valuable guidance.

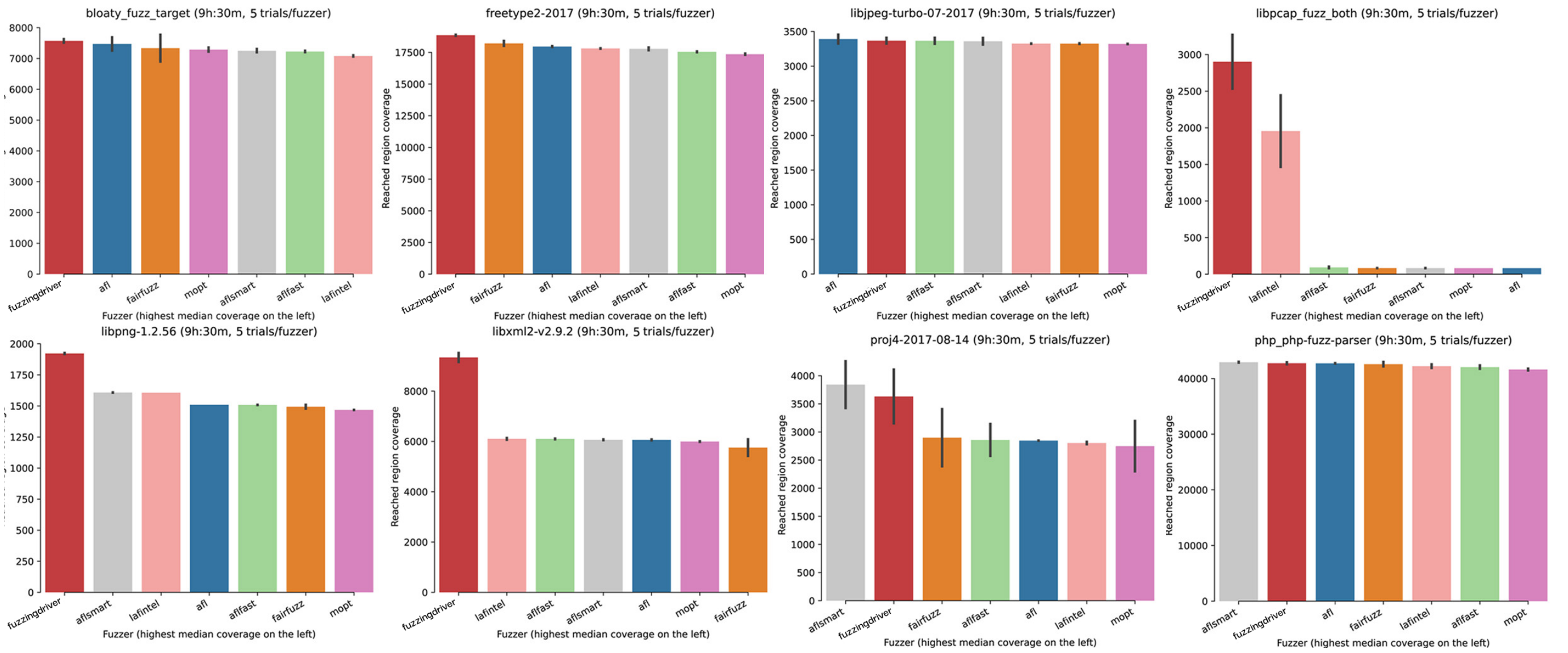

Fig. 1. The results of eight benchmarks on FuzzBench

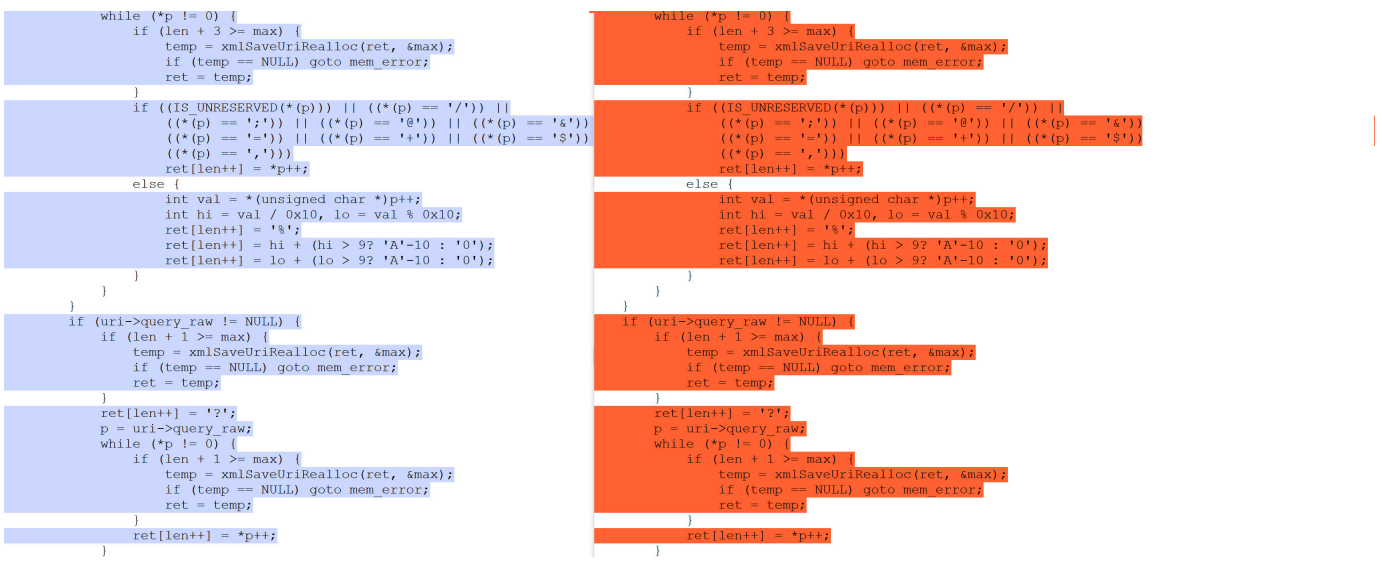

Fig. 2. Libxml code coverage, showing FuzzingDriver on the left and Google dictionaries on the right